\documentclass[sn-mathphys-num]{sn-jnl}

\usepackage{graphicx}%
\usepackage{multirow}%
\usepackage{amsmath,amssymb,amsfonts}%
\usepackage{amsthm}%
\usepackage{mathrsfs}%
\usepackage[title]{appendix}%
\usepackage{xcolor}%
\usepackage{textcomp}%
\usepackage{manyfoot}%
\usepackage{booktabs}%
\usepackage{algorithm}%
\usepackage{algorithmicx}%
\usepackage{algpseudocode}%
\usepackage{listings}%

\theoremstyle{thmstyleone}%
\theoremstyle{thmstyletwo}%

\theoremstyle{thmstylethree}%
\newcommand{\ket}[1]{\ensuremath{\left|#1\right\rangle}}

\begin{document}

\title[Article Title]{Experimental Simulation of Two Pulses and Three Pulses Coherent One Way Quantum Key Distribution Protocol in Noisy/Noiseless  and Wired/Wireless Environment}


\author*[1]{\fnm{Arijit} \sur{Roy}}\email{arijit21111998@gmail.com}

\author[1]{\fnm{Arpita} \sur{Maitra}}\email{arpita76b@gmail.com}

\author[2]{\fnm{Saibal} \sur{Kumar Pal}}\email{saibal.pal@gov.in}

\affil[1]{\orgdiv{Institute for Advancing Intelligence}, \orgname{TCG CREST}, \orgaddress{\street{Bengal Eco Intelligent Park Ln, EM Block, Sector V}, \city{Kolkata}, \postcode{700091}, \state{West Bengal}, \country{India}}}

\affil[2]{\orgdiv{Scientific Analysis Group (SAG)}, \orgname{Defence Research \& Development Organization}, \orgaddress{\street{MetCalfe House}, \city{New Delhi}, \postcode{110054}, \state{Delhi}, \country{India}}}


\abstract{Due to the rapid advancement of quantum technology, the traditional established classical cryptographic protocols are no longer secure. To make the world quantum safe, different quantum protocols have been taken into account. Quantum Key Distribution (QKD) protocols are one of them where two legitimate parties can securely communicate by upholding the quantum principles. Out of various QKD protocols, Coherent One Way (COW) protocol is one of the most famous protocol because of its ease of hardware deployment, and resilience nature  towards PNS attack. In this initiative, we have implemented the original version of two pulses COW QKD protocol and a very recent variant of it, three pulses COW (Phys. Rev. Applied 18, 064053 Published 19 December 2022), in Optisystem v21.1. We demonstrate the encoding as well as decoding portions of the protocols under both noisy and noiseless scenario considering different weather conditions. Finally, we report a comparative study amongst the protocols under wired (optical fibre) and wireless (free space) environments to check the proper integrity of transmission. The simulation results provide us an overview regarding the practical implementation of the protocols under different weather conditions.}

\keywords{Quantum Key Distribution, COW Protocol, Free Space Optics, Signal to Noise Ratio.}



\maketitle

\section{Introduction}\label{intro}
One of the most important aspects of contemporary telecommunications technology is the transfer of encrypted data from one side to another. Different cryptographic algorithms, such as RSA~\cite{RSA} and DH, have already been established in the current case scenario; however, those are vulnerable against Quantum Adversary due to Shor algorithm~\cite{shor}. This paved the way for Quantum Cryptography. Developing quantum encryption algorithms for safe data transfer is the need of the hours. 

In this direction, Quantum Key Distribution (QKD) protocols offer quantum-safe transmission. QKD protocols take benefits of quantum principles, such as i) the No Cloning Theorem, which states that no arbitrary quantum state can ever be completely duplicated, ii) the irreversibility of measurements, which indicates that an output state cannot be used to create the input state. Through this mode of communication, two authorized parties (Sender: Alice, Receiver: Bob) can exchange a secret private key that will be used to encrypt and decrypt any type of data, whether it be quantum or classical.

The first ever QKD protocol was proposed by Bennett and Brassard in 1984 named as BB84 protocol~\cite{BB84} where they used to encode the qubit by the polarization angles of single photon. The community paid close attention to this algorithm because of its unwavering security against various quantum attacks. Subsequently, other QKD methods were developed, some of which are theoretical~\cite{B92,98,th} and others of which are realized experimentally~\cite{COW1,COW2,COW3,COW4}. Different quantum groups have already established different QKD protocols such as Differential Phase Shift (DPS)~\cite{DPS}, Coherent One Way (COW)~\cite{COW3}, Measurement Device Independent (MDI)~\cite{MDI}, Twin Field (TF)~\cite{Twin field} QKD and so on. From the perspective of hardware implementation, the existence of inherent physical imperfection in the optical components creates the possibility of specialized attack methods, like side-channel attacks~\cite{side}. Exploiting the flaws in the single-photon detectors, and the weaknesses in the random number generators, a lot many attacks can be devised. Among all of them, the Photon Number Splitting (PNS) attack is a significant quantum attack that exploits the loopholes in the optical components. To resist this attacking strategy, the eavesdropper (let's say Eve) exploits the photon losses brought on by device flaws to retrieve shared key information. To neutralize the above mentioned attack Distributed Phase Reference (DPR) QKD protocols~\cite{DPR,DPR2} have entered into the picture. The two primary protocols that demonstrate resilience against PNS attacks are DPS and COW protocols and those are also very easily implementable in hardware. These schemes use very weak coherent signals containing mean number of photons $\mu$ ($\vert\alpha\vert^2$) $< 1$. Instead of encoding the data into only one single photon, these protocols are using to encode the information via phase (for DPS) and via time of arrival of photons (for COW). These encoding strategies withstand the PNS attack and keep the message's confidentiality.

As mentioned earlier, the COW QKD protocol is a part of the DPR protocol, which verifies security by the coherence of successive coherent pulses. However, there are certain flaws in the protocol itself. As the coherence, or more accurately, the security, is only examined for two or more consecutive coherent pulses, some attacks can be designed where without introducing any error, Eve can extract a significant amount of information about the raw key. In fact, in~\cite{zero,zero ppt,sequential}, the authors utilized a unique kind of intercept-resend attack called a sequential attack, in which Eve could have some information, modify it, and send it to Bob without making any errors~\cite{zero}. In such attack strategy, if Eve can get a conclusive measurement of two or more sequential coherent pulses and the measurement results start and end with vacuum pulses, then Eve can modify the whole message and send it to Bob without being noticed by the legitimate parties~\cite{zero ppt}, hence hampering the integrity of the protocol.

In order to close the aforementioned gaps in the protocol, Emilien Lavie~\cite{3 pulse} suggested various modifications to the original protocol in the motivation to make the protocol more secure. The modifications proposed in~\cite{3 pulse} are as follows;  a) adding one vacuum tail pulse to each of the states ($\vert\Phi_0\rangle$, $\vert\Phi_1\rangle$ and $\vert\Phi_d\rangle$); and b) utilizing a balanced (50:50) beam splitter at Bob's end for the passive basis measurement. By introducing these small changes, the protocol can resist the sequential attack.

In this work, we simulate both the protocols (2-pulses COW QKD~\cite{COW1} and 3-pulses COW QKD~\cite{3 pulse}) on Optisystem v21.1. in various scenarios. First, we consider the noiseless scenario followed by a noisy scenario. We also simulate the protocols in wired and wireless environment.

The manuscript is arranged as flows;  first we describe the COW QKD protocol in Sec.~\ref{description},  Sec.~\ref{cow} talks about the simulation models at every node of the layout, Sec.~\ref{results} tells about the results of simulation and also discusses the comparative analysis of the protocols for long distance communication and in Sec.~\ref{conclusion}, we finally conclude the manuscript providing the future research insight.\\
\section{Description of COW QKD Protocol:}\label{description} In this section, we describe the original 2-pulses COW QKD protocol. The protocol is enumerated below.
\subsection{2-Pulses COW QKD}
\begin{enumerate}
    \item Alice begins with a weak coherent pulse generator with an average mean number of photons per pulse $\vert\alpha\vert^2 = \mu  < 1$. She encodes the cipher text by the combination of the coherent and empty pulses. The logical bits are encoded as follows; $0_L \rightarrow \vert\Phi_0\rangle = \vert0\rangle\vert\alpha\rangle$, $1_L \rightarrow \vert\Phi_1\rangle = \vert\alpha\rangle\vert0\rangle$ and Decoy $\rightarrow \vert\Phi_d\rangle = \vert\alpha\rangle \vert\alpha\rangle$ with probabilities of $\frac{1-f}{2}$ for $\vert\Phi_0\rangle$ , $\vert\Phi_1\rangle$ and probability $f$ for $\vert\Phi_d\rangle$ states. The separation between two pulses is a pre-determined value of $\Delta t$.
    \item The entire encoding pulse train is now transmitted to Bob over the quantum channel which is having a transmission coefficient $t=10^{-\alpha_{att}l/10}$; where $\alpha_{att}$, and $l$ are the attenuation coefficient and length of the fibre respectively.
    
    \begin{figure}[h]
    \begin{center}
    \includegraphics[width = 11 cm, height = 4 cm]{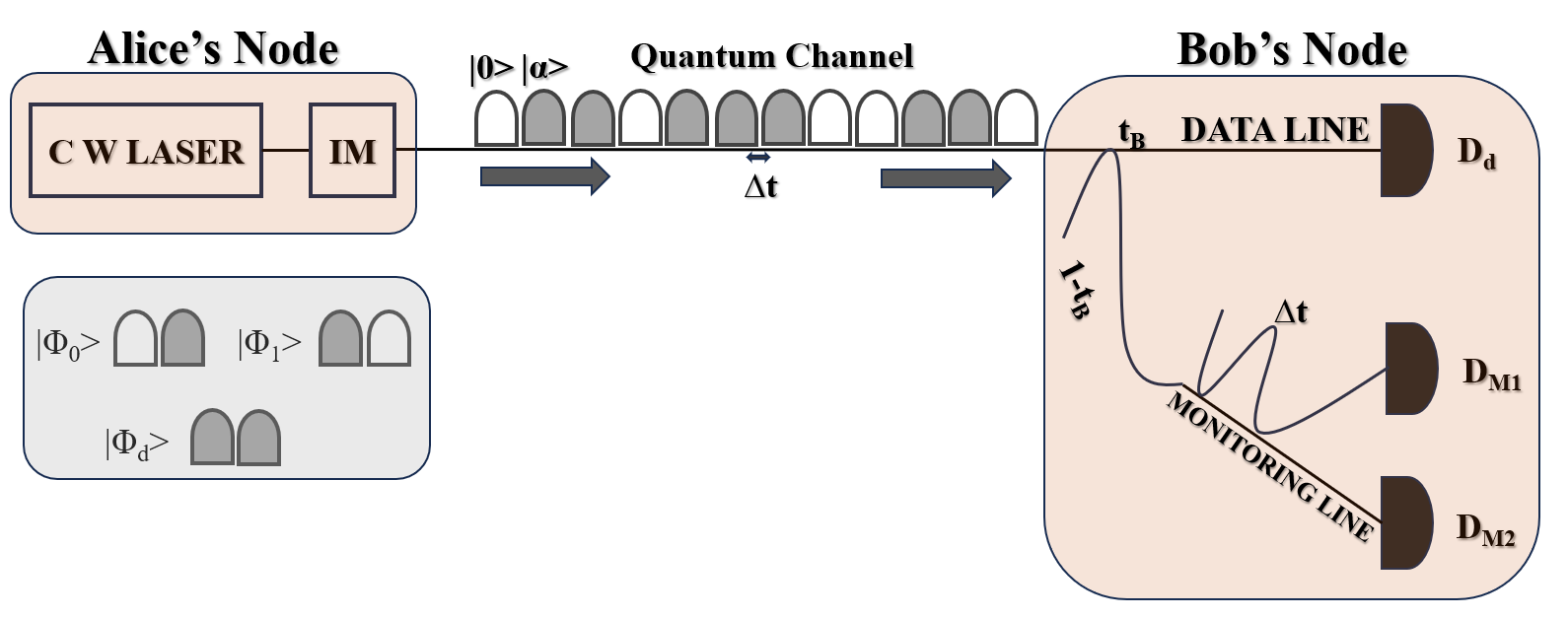}
    \caption{Schematic of COW QKD Protocol.}\label{schematic}
    \end{center}
\end{figure} 

 \item Bob measures 90\% of the photons at Data Line for the purpose of creating the raw key, with the remaining 10\% being used to verify the transmission's security at the Monitoring Line [ Fig.~\ref{schematic}].
    \item At Data Line only one photo-detector ($D_d$) is there which unambiguously discriminate the logical bits via the time stamps of the incoming photons. Precisely, Bob assigns the bit value $0_L$ ( $\ket{\Phi}_0$) if the ``Click" in $D_d$ in the first time instant and if the ``Click" is in second time instant, then it would consider as bit value $1_L$ ( $\ket{\Phi}_1$). But there is a confusion between two consecutive non-empty pulses, because it is very difficult to discriminate the  $\ket{\Phi}_2$ state from the two consecutive  $\ket{\Phi}_0$
 and  $\ket{\Phi}_1$ states. To mitigate such error, Bob publicly announces over an authenticated classical channel in which two pulse sequences he got detection at $D_d$, but he does not reveal the particular time slots of ``Click''ing. If the pulses belongs to decoy states, Alice says to throw away those states; otherwise accept the sequence of pulses.
 \item The monitoring line is used to get the knowledge of the presence of eavesdropping. Here, the clicking of the detector $D_{M_1}$ and  $D_{M_2}$ for the non-consecutive pulses are random. In the path of $D_{M_1}$, we introduce a time delay $\Delta t$. The interferometer is arranged in such a way so that the adjacent coherence pulses $\ket{\alpha}$ could not produce a ``Click" at $D_{M_2}$, due to the time delay $\Delta t$ at $D_{M_1}$, the interference is happen only at the path of $D_{M_1}$. To quantify the proper coherence between the consecutive non-empty pulses we should check the Visibility $V_s$ which is defined as; 
\begin{eqnarray*} 
    V_s = \frac{P_{click}(D_{M_1}|_s) - P_{click}(D_{M_2}|_s)}{P_{click}(D_{M_1}|_s) + P_{click}(D_{M_2}|_s)}
\end{eqnarray*}
where $s \in S = \{\text{decoy}, 0_L 1_L, 0_L \text{decoy}, \text{decoy} 1_L, \text{decoy decoy}\}$, i.e., $S$ contains all the possible signal combinations that have two or more adjacent coherent pulses $\vert\alpha\rangle$. Here, $P_{click}(D_{M_i}|s)$, $i=\{1,2\}$ is the conditional probability that detector $D_{M_i}$ clicks from the above mentioned $s$ values. The values of $s$ are selected in such a way so that, there should be two or more consecutive successive non-empty pulses.
For ideal case,  due to the time delay ($\Delta t$), $D_{M_2}$ should never click for the above combinations of $S$; so giving an unit value of Visibility ($V_s = 1$). Any fall in Visibility indicates the presence of Eve, the adversary.
\item Finally, the sifting procedure (removing decoy pulses and disclosing some actual bits) is done and the Quantum Bit Error Rate (QBER) is estimated. This should be less than a pre-defined threshold value. Finally, we apply error correction and information reconciliation on the sifted key to obtain the secret key.
\end{enumerate}
\subsection{3-Pulses COW QKD}
In 2-pulses COW QKD, the security is only being checked for two or more consecutive coherent pulses, whereas if there are two or more consecutive vacuum pulses at the pulse train, then that may be captured by the Eve, which opens a new avenue of the attacks. In~\cite{sequential}, the authors showed that the loophole which is created by the consecutive vacuum pulses enables sequential attack to be successful. They demonstrated that if Alice's string or some of the portions of the string of quantum states starts (start) and ends (end) with vacuum pluses, then the total information is being leaked to Eve. So, the authors concluded that the vacuum pulses are hampering the effectiveness of COW protocol.

To mitigate the above mentioned loophole, 3-pulses COW QKD protocol has been developed~\cite{3 pulse}. In this work, the authors proposed slight changes in the original protocol. They proposed i) to incorporate an additional vacuum tail pulse at each of the states and ii) provide one balanced (50:50) beam splitter at Bob's end. 
\ \\
\ \\
 \textbf{Reasons for the above changes :}
\begin{enumerate}
    \item By incorporating one vacuum pulse at the end of each state does not allow to produce a lot of consecutive coherent pulses, only the consecutive coherent pulses can occur at decoy state and only this decoy state will further be used for security checking.
    \item In the original protocol, the authors were using 9:1 BS, resulting a very less amount of photons passing through the monitoring line. However, that was not a big issue as there were already a lot of cases for consecutive coherent pulses ($0_L1_L, \text{decoy} 1_L, \text{decoy} \text{decoy}$ etc.) which can be used for the security checking purpose. Contrary to this, by adding an additional vacuum pulse as a tail to each pulse, only one case of consecutive coherent pulses can occur which is actually the $\text{decoy}$ state. Hence, to check the proper integrity of the security of the protocol, we need to send a very large bit sequence at the monitoring line to geta large number of decoy states. In order to ensure that Visibility can be determined with certainty, it is therefore preferable to utilize a balanced BS at Bob's end such that a significant number of decoy states will reach the monitoring line without utilizing a huge bit sequence. 
\end{enumerate}

\section{Simulation of  COW QKD protocol :}\label{cow}
In this section,  we are going to present the simulation model~\cite{simulation cow} of 2-pulses and 3-pulses versions of COW QKD protocol. From the simulation results, we further check the maximum possible achievable distance of transmission with maintaining proper integrity of the protocol.
\subsection{2-Pulses COW QKD :}\label{2 pulse}
\begin{figure}[h] 
    \centering
    \includegraphics[width = 13 cm, height = 6 cm]{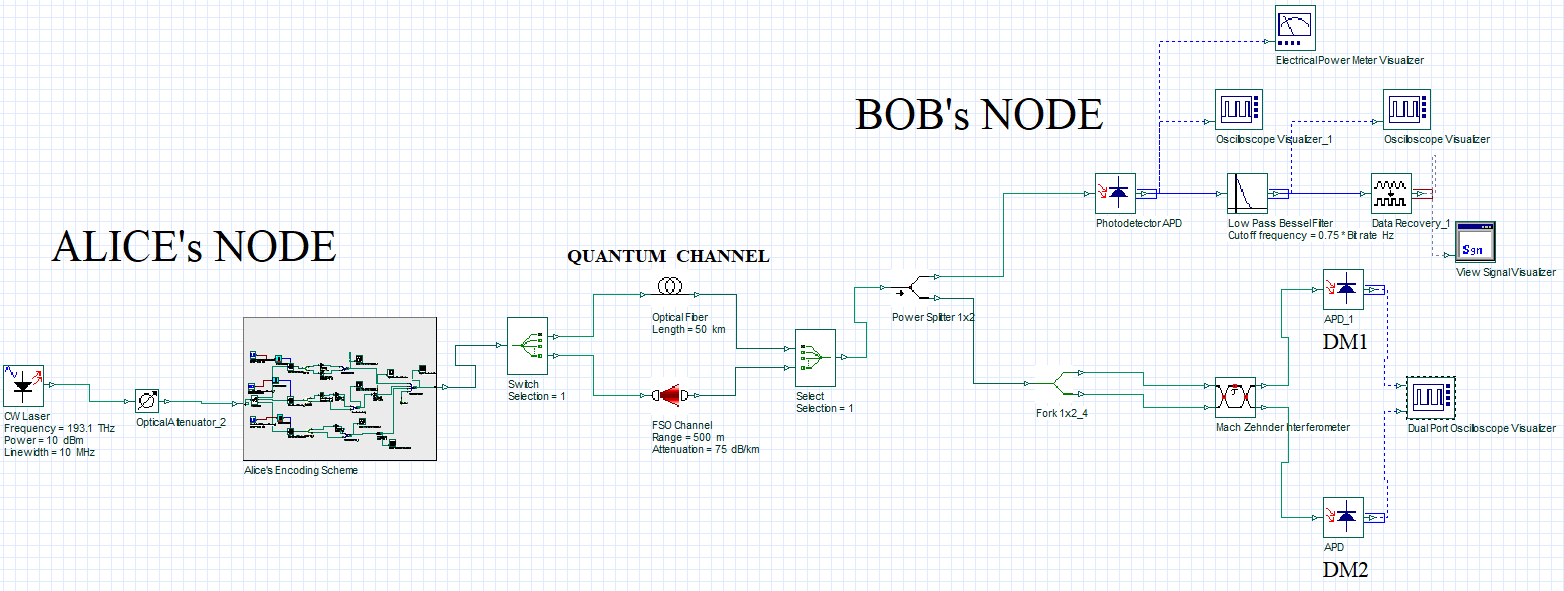}
    \caption{Simulation Design of COW QKD Protocol.}\label{simulation}
\end{figure}

In this portion of our report, we are going to describe the configuration of the simulation model of the 2-pulses COW protocol which is depicted in Fig.~\ref{simulation}. There are basically three partitions in our model; a) Alice's Node (Qubit generation and Encoding node), b) Quantum Channel (Transmission medium of encoded qubits), c) Bob's Node (Decoding and security checking node). The designed model and all the output results have been measured at a data rate of $1$ Gbps which is sufficient for long distance communication.
\subsubsection{Alice's Node:}\label{A}
\begin{figure}[h] 
    \centering
    \includegraphics[width = 13 cm, height = 5.1 cm]{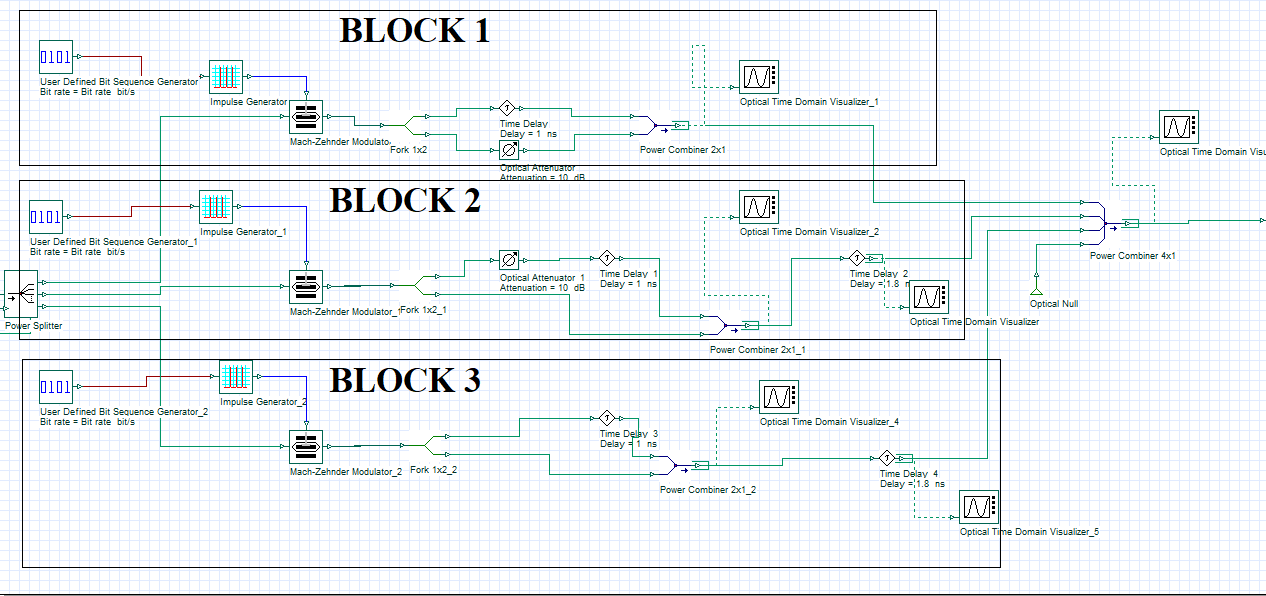}
    \caption{Layout Configuration of ALICE's NODE.}\label{alice2}
\end{figure}

In the initial phase of the experimental set up of Alice's side [Fig.~\ref{alice2}], a Continuous Wave (CW) Laser is characterized by a central frequency of $193.1$ THz (optical wavelength $1552.52$ nm) because of its high power efficiency and a low linewidth of $10$ MHz containing signal power of $10$ dBm which is joined with an Optical Attenuator (OA) in order to make the optical signal very weak less than unity to neutralize the effect of PNS attack.\\
\\
Now, the weak signal is passing through the Select Component working as Beam Splitter (BS) into three distinct encoding circuits; those are described in the following manner.

\begin{enumerate}
    \item {\textbf{Encoding circuit for $\vert0\rangle$$\vert\alpha\rangle$ state we use BLOCK 1 circuit.}}
    This circuit starts with a pre-defined Bit Sequence Impulse Generator for its low FWHM (Full Width Half Maximum) value showcasing less decoherence and less dispersion effect during the transmission. With a proper input bit sequence, the electrical impulses are generated and get connected with the optical signal through the Mach Zhender Modulator (MZM). The output of MZM will lead to give a Weak Pulsed Coherent signal. Now, this signal is split by a Fork $1\times2$ component for the modification in intensity/power level [Appendix] of the signal for encoding purpose. To make the state $\vert0\rangle$$\vert\alpha\rangle$, one of the splitted signal is passing through the OA with attenuation $10$ dB and another splitted signal is passing through a time delay component possessing a time delay $1$ ns without any attenuation. At last those two signals are merged in a Power combiner $2\times1$ and finally gives us our required state which is visualized by an Optical Time Domain Visualizer [Fig.~\ref{alpha alpha}(a)]
    
    \item {\textbf{Encoding circuit for $\vert\alpha\rangle$$\vert0\rangle$ state we use BLOCK 2 circuit.}}
The circuit is initially the same as the previous one. The only modifications are at the encoding level, where one of the splitted pulsed signal is going through the OA as well as the time delay component, with the parameter previously specified. The other signal is not impacted in any way. At a power combiner, the delayed and the attenuated pulse combine with the unattenuated pulse. We add a time delay component to the combiner's output to ensure correct time synchronization with the other quantum states. Using the visualizer, we can finally view our intended state that is correctly time-synchronized [Fig.~\ref{alpha alpha}(b)].

    \item{\textbf{Encoding circuit for $\vert\alpha\rangle$$\vert\alpha\rangle$ state we use BLOCK 3 circuit.}}
In this instance, the initial phase is the same as the previous one; the only difference is that one pulse experiences a temporal delay during the encoding phase before they are combined at the combiner. For correct time synchronization, there is an additional time delay component [Fig.~\ref{alpha alpha}(c)] at the end of the combiner.
\end{enumerate}
The output states of the circuits mentioned above are finally combined in a power combiner and the final encoding string of Alice is being visualized by an Optical Time Domain Visualizer which is given in [Fig.~\ref{alice string}].
\subsubsection{Quantum Channel:}\label{QC}
We select Free Space Optics (FSO) and Optical Fibre (OF) as the channels for qubit transmission; these can be substituted with our requirements by the combination Switch and Select components. The reference wavelength for OF is $1550$ nm, and the attenuation in the fibre is $0.2$ dB/Km, additionally, the signal exhibits self-phase modulation. For FSO, the attenuation values vary depending upon the weather conditions, as shown in TABLE~\ref{tabble1}.

\begin{table}[h]
\caption{Attenuations at different atmospheric conditions~~\cite{FSO}}\label{tabble1}%
\begin{tabular}{@{}llll@{}}
\toprule
\textbf{Atmospheric phenomena} & \textbf{Attenuation [dB/Km]}\\
\midrule
\hspace{1.2cm}Very Clear    & \hspace{1.2cm}0.4\\
\hspace{1.2cm}Clear         & \hspace{1.2cm}0.57\\
\hspace{1.2cm}Light Haze    & \hspace{1.2cm}1.5\\
\hspace{1.2cm}Haze          & \hspace{1.2cm}3.8\\ 
\hspace{1.2cm}Light Rain    & \hspace{1.2cm}6.27\\
\hspace{1.2cm}Moderate Rain & \hspace{1.2cm}9.64\\
\hspace{1.2cm}Heavy Rain    & \hspace{1.2cm}19.28\\
\hspace{1.2cm}Light Fog     & \hspace{1.2cm}18\\
\hspace{1.2cm}Moderate Fog  & \hspace{1.2cm}28.9\\
\hspace{1.2cm}Heavy Fog     & \hspace{1.2cm}75\\
\botrule
\end{tabular}
\end{table}
\subsubsection{Bob's Node:}\label{B}
\begin{figure}[h] 
    \centering
    \includegraphics[width = 12 cm, height = 6 cm]{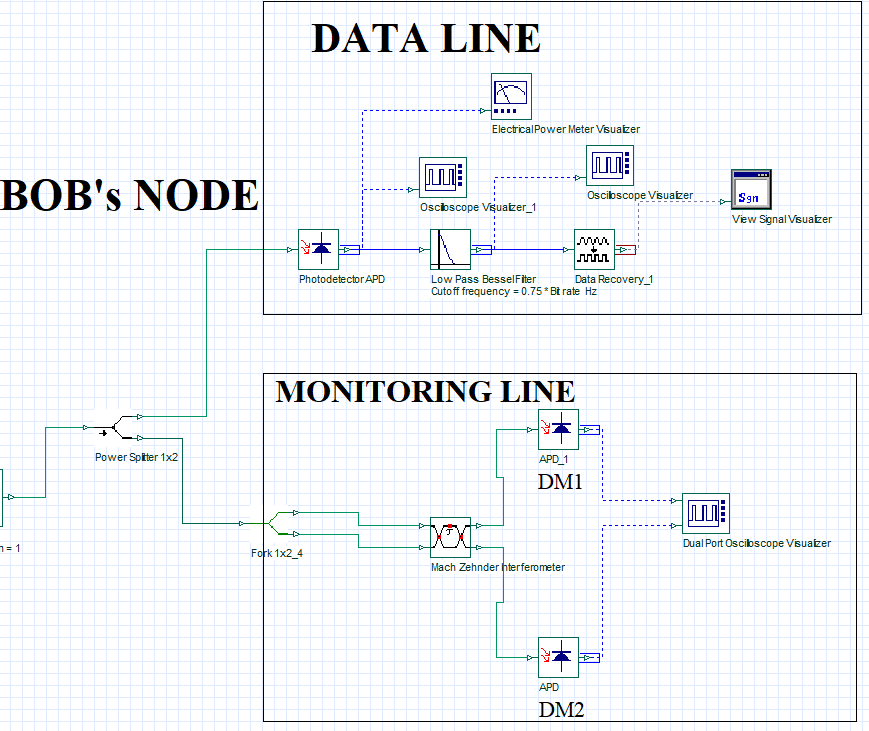}
    \caption{Simulation Model at BOB's end.}\label{bob}
\end{figure}

Bob's circuit configuration [Fig.~\ref{bob}] starts with a power splitter with a splitting ratio of 9:1; majority of the signal has been gone to the Data line which is used as decoding the qubits and the residual signal goes to the Monitoring line which is being used for security checking purpose.\\
\begin{itemize}
 \item $\textbf{Data Line :}$
At this stage of layout, only a Avalanche Photodiode (APD) is placed to detect and to measure the time stamps of the incoming photons (stipulated parameters of APD are given in TABLE ~\ref{table2}). We study both Noiseless [Fig.~\ref{noise}(a)] and Noisy [Fig.~\ref{noise}(b)] scenarios of APD. We take a) Signal to ASE noise, b) ASE to ASE Noise and c) Thermal Noise only (Noise values mentioned in TABLE~\ref{table2}). The Oscilloscope Visualizer is utilized for the purpose of visualization, while the Electrical Power Meter is employed to measure the signal's power. The Signal to Noise Ratio (SNR) of the model will be determined using this information.\\

\item  $\textbf{Monitoring Line :}$
This portion of the layout begins with a Mach Zhender Interferometer (MZI) in one of its arms having a time delay ($\Delta$t) equal to the exact same amount of time that separates two pulses at Alice's side, i.e.,  $1$ ns. At the output end of the MZI we place two APDs and named as DM$_1$ (with time delay of $1$ ns) and DM$_2$ (without delay).\\
\end{itemize}
One should note that all the parameters' values at Alice's node, Bob's node and for the quantum channel are reported in TABLE~\ref{table2}.
\begin{table}
\caption{Specified parameters at various layout components.}\label{table2}%
\begin{tabular}{@{}llll@{}}
\toprule
\textbf{Components} & \textbf{Parameters} & \textbf{Stipulated Values}\\
\toprule
C W LASER     & Freq./Wavelength ($\lambda$) & 193.1 THz/1552.5 nm\\
              & Linewidth & 10 MHz\\
\toprule
             &  \hspace{0.15cm}\textbf{\large{Alice's Node}}\\
\toprule
Mach-Zhender Modulator (MZM)        & Extinction Ratio & 30 dBm\\
Optical Attenuator         & Attenuation ($\alpha$) & 10 dBm\\ 
Time delay Component    & Delay ($\Delta$t) & 1 ns\\
\toprule
             & \textbf{\large{Quantum Channel}}\\
\toprule
Optical fibre (OF) & Reference Wavelength & 1550 nm\\
\hspace{1.2cm}   & Attenuation & 0.2 dB/Km\\
\hspace{1.2cm}     & Dispersion & 16.75 ps/nm.Km\\
\hspace{1.2cm}  & Dispersion Slope & 0.075 ps/nm$^2$.Km\\
\hspace{1.2cm}     & Self Phase Modulation & Enable\\
Free Space (FSO)     & Transmitter Aperture Diameter (d$_t$) & 5 cm\\
\hspace{1.2cm}     & Receiver Aperture Diameter (d$_r$) & 20 cm\\
\hspace{1.2cm}    & Beam Divergence (D) & 2 mrad\\
\hspace{1.2cm}    & Attenuation ($\alpha$) & from TABLE \ref{tabble1}\\
\toprule
             &  \hspace{0.15cm}\textbf{\large{Bob's Node}}\\
\toprule
Avalench Photodiode (APD) & Gain & 30\\   
      \hspace{1.2cm} & Responsivity & 1 A/W\\  
      \hspace{1.2cm} & Dark Current & 10 nA\\  
      \hspace{1.2cm} & Signal to ASE Noise & Enable\\  
      \hspace{1.2cm} & ASE to ASE Noise & Enable\\  
      \hspace{1.2cm} & Thermal Noise & 10$\times$10$^{-27}$ W/Hz\\
Low pass Bessel Filter & Cutoff Freqequency & 0.75$\times$Bit Rate\\
Mach-Zhender Interferometer (MZI) & Time delay ($\Delta$t) & 1 ns\\ 
\botrule
\end{tabular}
\end{table}
\subsection{3-Pulses COW QKD :}\label{3 pulse}
Higher security is demonstrated by the 3-pulses COW QKD~\cite{3 pulse}. By modifying the current simulation layout with a few minor adjustments suggested by Emilien Lavie~\cite{3 pulse} at Bob's and Alice's nodes, we simulate the 3-pulses COW QKD protocol.
\begin{figure}[hbt!]
    \centering
     \includegraphics[width = 10 cm, height = 4.2 cm]{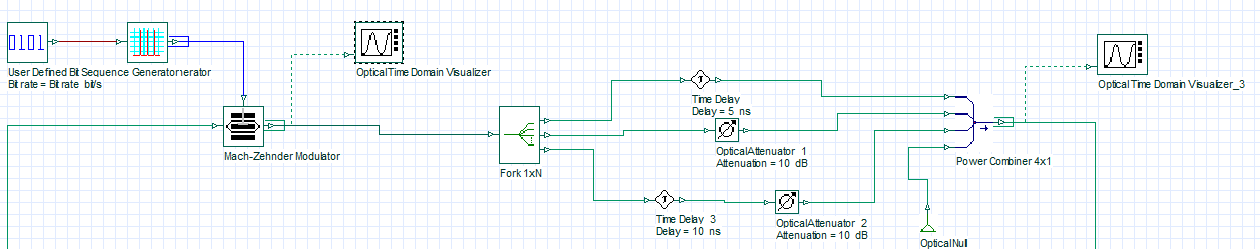}
    \caption{Modification at Alice's Node.}\label{alice3}
\end{figure}

 The modifications are as follows. We selected a pulse interval of $5$ ns at each block at Alice's end, and we added an extra delayed path with a $10$ dB attenuation for introducing one more vacuum pulse in each BLOCK. Here, we  present only the layout for BLOCK 1 [Fig.~\ref{alice3}]  as the modifications remain the same for all the blocks. The decoy state circuit has $3$ dB of attenuation to distinguish the state at Bob's monitoring line segment.
 For the passive selection of the two lines (Data and Monitoring) at Bob's end, we just modify the splitting ratio from 9:1 to 1:1 in the BS as suggested in~\cite{3 pulse}. \\
\section{Simulation Results and Corresponding Interpretation:}\label{results}
In this section, we provide all the simulation results with proper clarification. The observations at different nodes, i.e., at Alice's end and at Bob's end for both the protocols are presented below.
\subsection{Simulation results at Alice's end:}\label{results alice}
\begin{itemize}
\item $\textbf{Results for 2-Pulses Version :}$
We know that the protocol exploits three quantum states i) $\vert0\rangle$$\vert\alpha\rangle$, ii) $\vert\alpha\rangle$$\vert0\rangle$ and iii) $\vert\alpha\rangle\vert\alpha\rangle$. In the software, the states are generated using BLOCK 1, BLOCK 2, and BLOCK 3 respectively. In the visualizer, the representation of the states are presented in Fig~\ref{alpha alpha}.  Visualizer\_1 shows the state $\vert0\rangle$$\vert\alpha\rangle$, Visualizer\_2 represents the state $\vert\alpha\rangle$$\vert0\rangle$, and the Visualizer\_4 reflects the state $\vert\alpha\rangle\vert\alpha\rangle$. In  Visualizer\_3 (Fig~\ref{alice string}), finally we can see the time synchronized entire pattern which has been sent to Bob from Alice. Here, the pattern we consider is $\{0_L,1_L,\text{decoy},1_L,0_L,1_L\}$.

\begin{figure}[h]
    \centering
    \includegraphics[width = 6 cm, height = 5 cm]{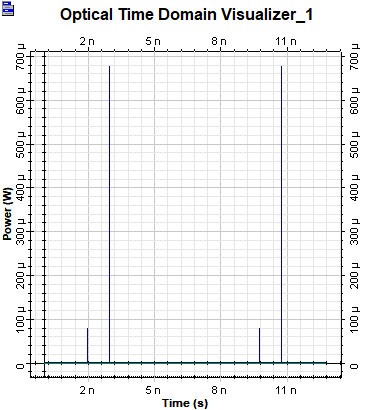}
    \includegraphics[width = 6 cm, height = 5 cm]{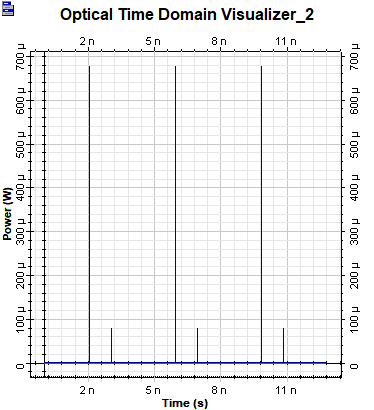}\\
    (a)$\hspace{6cm}$(b)\\
    \end{figure} 
\begin{figure}[h]
    \centering
    \includegraphics[width = 6.5 cm, height = 4.9 cm]{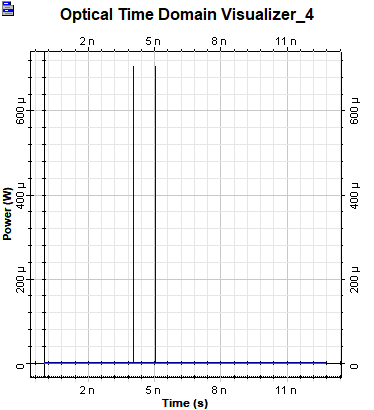}\\
    (c)\\   
    \caption{Different quantum states at Alice's Node (a) $\vert0\rangle$$\vert\alpha\rangle$, (b) $\vert\alpha\rangle$$\vert0\rangle$ and (c) $\vert\alpha\rangle$$\vert\alpha\rangle$ state.}\label{alpha alpha}
\end{figure}
\begin{figure}[hbt!] 
    \centering
    \includegraphics[width = 10 cm, height = 5.4 cm]{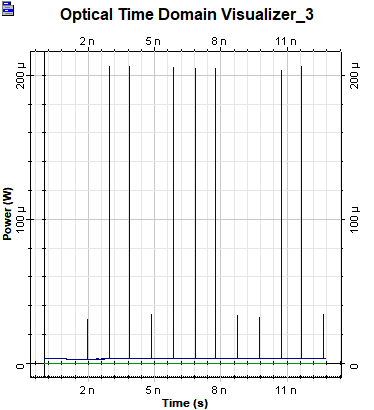}
    \caption{Encoded String of Alice after time synchronization.}\label{alice string}
\end{figure}

\item  $\textbf{Results for 3-Pulses Version :}$\\
In the Fig.~\ref{alice string 3}, the encoded signal is presented by three sequential pulses where every quantum state starts and finishes with lower amplitude pulses and for the decoy states the amplitude ($\beta$) is less than the amplitude of the coherence pulse ($\alpha$), i.e., $\beta < \alpha$. Here, the pattern which is transmitted to Bob is $\{0_L,1_L,\text{decoy},1_L,\text{decoy},0_L\}$.
\begin{figure}[hbt!]
    \centering
    \includegraphics[width = 10 cm, height = 4.8 cm]{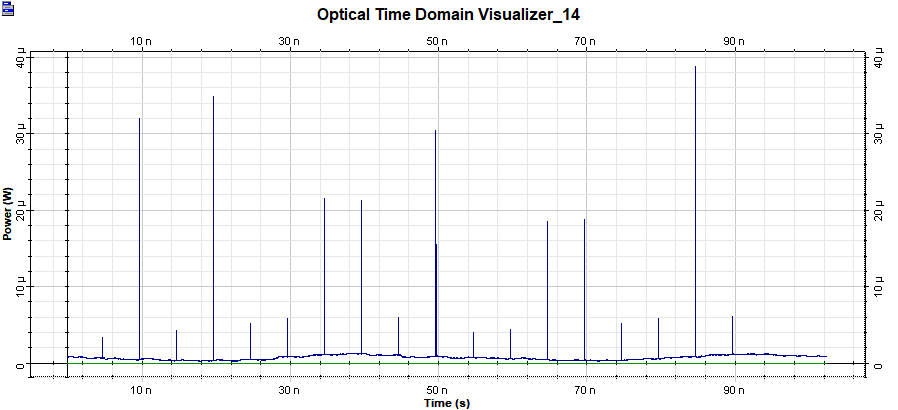}
    \caption{Encoding string of Alice for 3-pulses COW QKD.}\label{alice string 3}
\end{figure}
\end{itemize}
\subsection{Simulation results at Bob's end :}\label{results bob}
At Bob's side there are two nodes i) Data Line and ii) Monitoring Line which are used for the decoding of the quantum states and security checking of the protocol, respectively.
\begin{figure}[hbt!] 
    \centering
    \includegraphics[width = 6 cm, height = 5 cm]{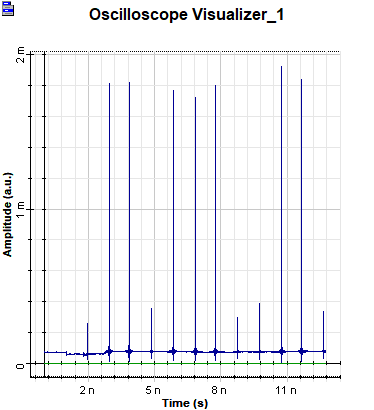}
     \includegraphics[width = 6 cm, height = 5 cm]{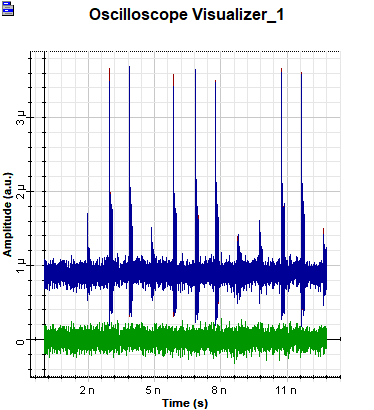}\\
     (a)$\hspace{6cm}$(b)\\
    \caption{Output of Data line (a) without noise and (b) with noise. for 2-pulses COW QKD}\label{noise}
\end{figure}
\begin{figure}[hbt!] 
    \centering
    \includegraphics[width = 6 cm, height = 5 cm]{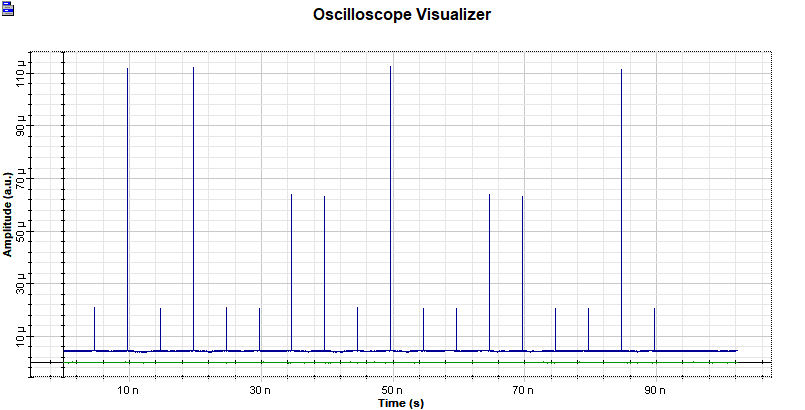}
     \includegraphics[width = 6 cm, height = 5 cm]{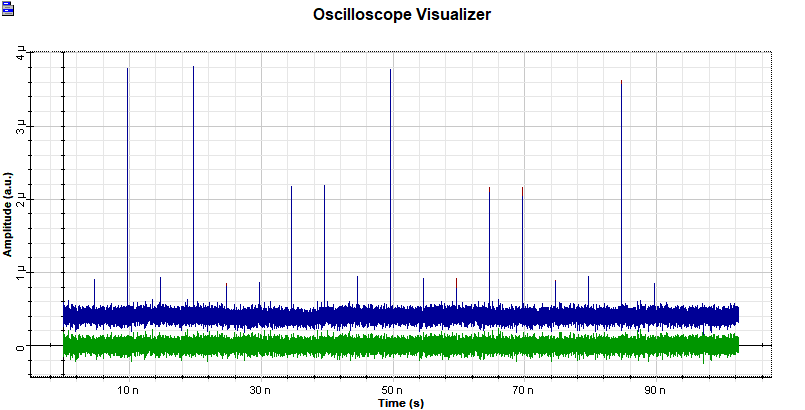}\\
     (a)$\hspace{6cm}$(b)\\
    \caption{Output of Data line (a) without noise and (b) with noise. for 3-pulses COW QKD}\label{noise2}
\end{figure}

\begin{figure}[hbt!] 
    \centering
     \includegraphics[width = 6 cm, height = 5 cm]{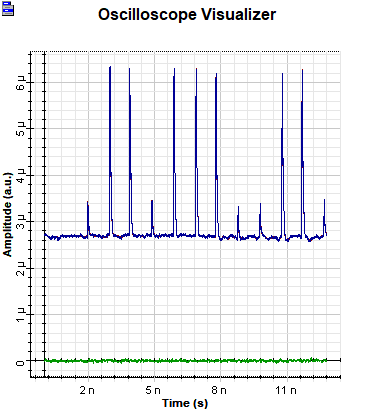}
     \includegraphics[width = 6 cm, height = 5 cm]{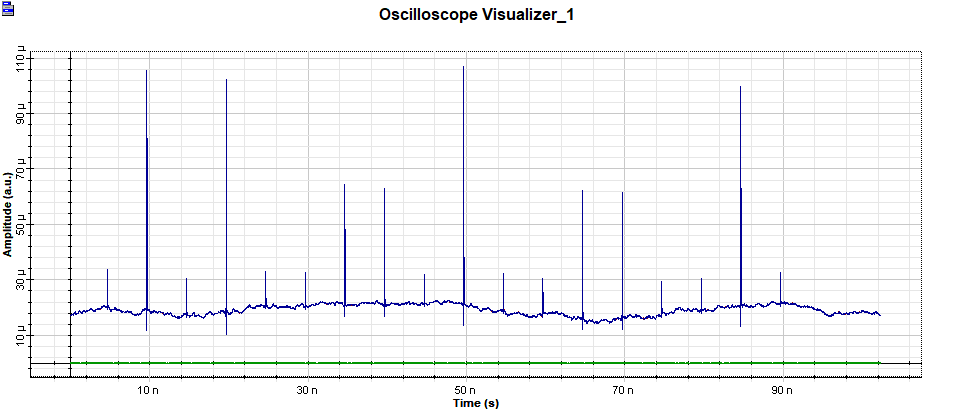}\\
     (a)$\hspace{6cm}$(b)\\
    \caption{Pulse train after filtration (a) for 2-pulses COW (b) for 3-pulses COW}\label{filter}
\end{figure}

\begin{figure}[hbt!] 
    \centering
    \includegraphics[width = 12 cm, height = 4 cm]{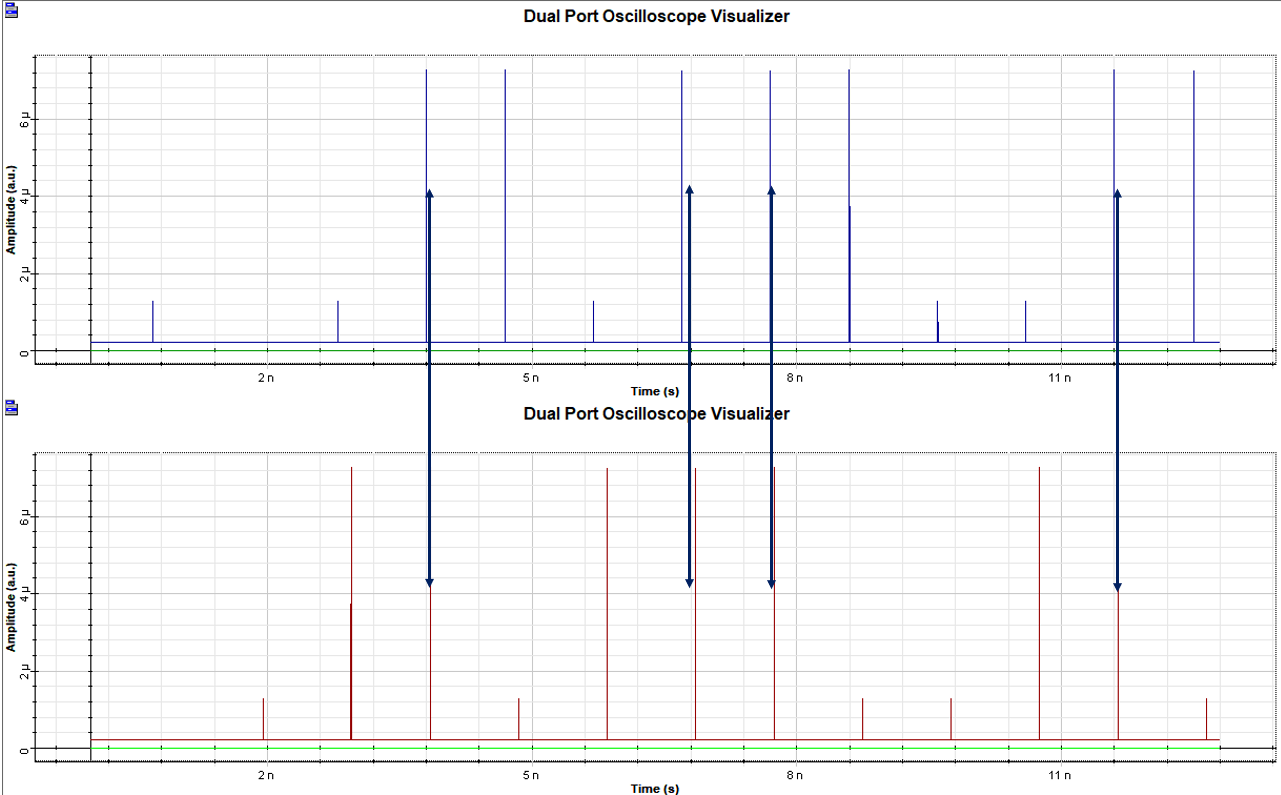}\\
    (a) For 2 - Pulse COW QKD\\
    \includegraphics[width = 12 cm, height = 4 cm]{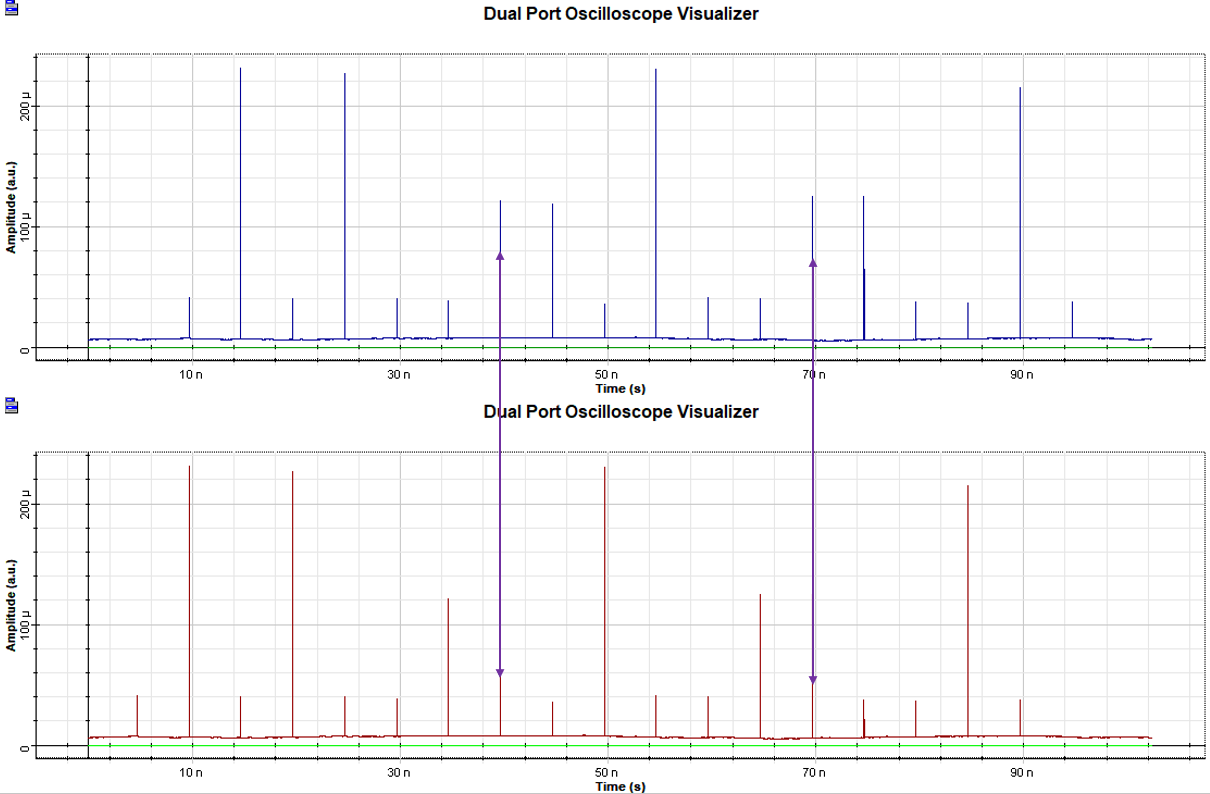}\\
    (b) For 3 - Pulse COW QKD\\
    \caption{Single view at Monitoring line.}\label{single mL}
\end{figure}

\begin{figure}[htp] 
    \centering
    \includegraphics[width = 12 cm, height = 3.3 cm]{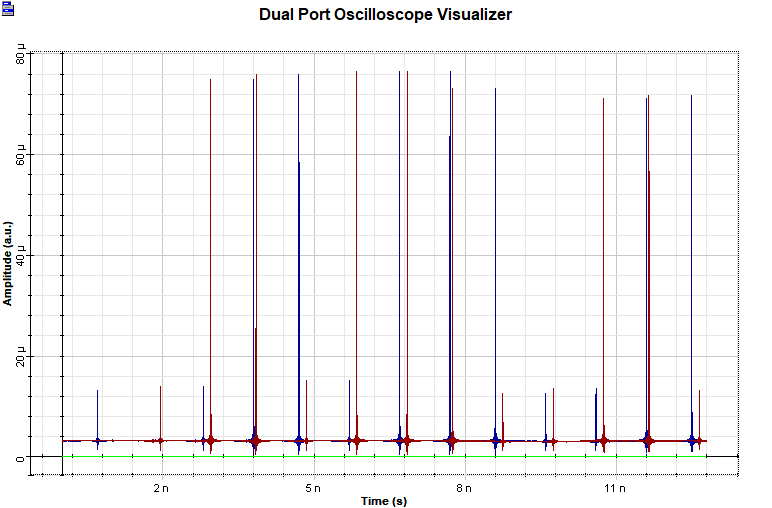}\\
    (a) For 2 - Pulse COW QKD\\
    \includegraphics[width = 12 cm, height = 3.3 cm]{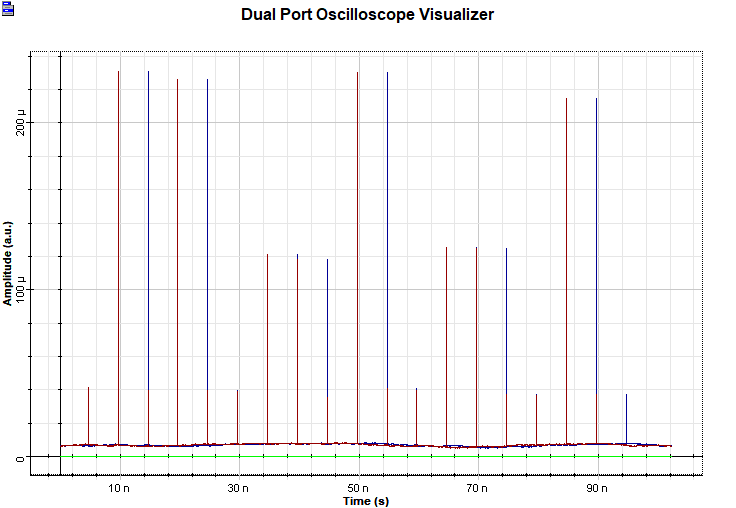}\\
    (b) For 3 - Pulse COW QKD\\
    \caption{Double view at Monitoring line.}\label{double mL}
\end{figure}

In Fig.~\ref{noise}(a) and Fig.~\ref{noise2}(a), the signal received by Bob at the Data Line is presented. Note that these are same as the pattern chosen by Alice in 2-pulses COW [Fig.~\ref{alice string}] and 3-pulses COW [Fig.~\ref{alice string 3}] respectively. Here, we assume that no signal is diverted to the Monitoring line. Then, we introduce some noise (noise parameters are mentioned at TABLE~\ref{table2}) in the data line. In Fig.~\ref{noise}(b) and Fig.~\ref{noise2}(b), the green portion denotes the signal's noise. A First Order Low Pass Bessel Filter with a $750$ MHz cutoff frequency is used to eliminate the associated noise with the original signal. The output results appear in Fig.~\ref{filter} after the filter. Even though the signal is somewhat distorted but it is still distinguished from the noise.
\\
Now, we consider the Monitoring Line. Here, we assumed that no signal is passed through the Data Line. The original pulse train and the $1$-bit delayed pulse train interfere at the monitoring line, and the security is only tested considering the successively consecutive higher pulses. We may check for the interference in the higher pulses by examining the output result [Fig.~\ref{single mL}] of monitoring line (blue, which is a delayed signal, interfering with the red, which is a non-delayed signal). The first lower pulse of the delayed signal (blue one) [Fig.~\ref{single mL}(a)] should be neglected since it is the last pulse of the original signal which is repeating because of 1 bit delay, so, the counting of pulses will start from the $2^{nd}$ lower pulse. Here, in Fig.~\ref{single mL}(a) $2^{nd}$ pulse of blue one (delayed one) interferes with the $3^{rd}$ pulse of the red one (non-delayed one); similarly pulse numbers $5$, $6$ and $10$ of the blue one (delayed) are interfering with the pulse numbers $6$, $7$ and $11$ of the red one (non-delayed) respectively. In Fig.~\ref{single mL}(b) $7^{th}$ pulse of the blue one (delayed) is interfering with $8^{th}$ number pulse of the red one (non-delayed) and $13^{th}$ number pulse of blue one (delayed)is interfere with the $14^{th}$ pulse of the red one (non-delayed). In Fig.~\ref{single mL}(b) it is clearly visible that the higher pulse interference only occur for the decoy states, any other combinations of states have not occurred at all. The consecutive interfering higher pulses are marked in Fig.~\ref{single mL}(a) and Fig.~\ref{single mL}(b). As we do not consider any eavesdropping in our simulation model, we are not getting any disruption at the interference level for both the protocols. One can also check the double view pattern of the monitoring line for their better understanding [Fig.~\ref{double mL}]. In Fig.~\ref{double mL}(a) the pulses positioned at (counting starts from the $2^{nd}$ pulse, neglecting the $1^{st}$ lower pulse (discussed earlier)) $3$, $6$, $7$ and $11$ are interfering signals whereas in Fig.~\ref{double mL}(b) pulse positioned at $8$ and $14$ are the interfering signals. 
\\
\\
\subsection{Signal to Noise Ratio (SNR) Calculation : }\label{SNR}
In order to verify the practical implementation of the protocol, there are a number of things we need to look into and check when determining whether a long-distance communication technique is compatible. The COW protocol uses optical signal amplitudes to discriminate between different states ($\vert0\rangle$ and $\vert\alpha\rangle$); if any of these amplitudes becomes equal because of attenuation at channel, atmospheric turbulence at free space, or Eve's presence, the protocol's overall integrity will be compromised. Signal to Noise Ratio (SNR) testing ought to be a crucial task for the checking of proper transmission of signals. \\
\begin{eqnarray*}
    SNR (\text{in dB}) = 10\log_{10} \frac{P_{signal}}{P_{noise}}
\end{eqnarray*}
where $P_{signal}$, $P_{noise}$ are the power values of the signal and induced noise.

Here, we carry out a comparative analysis of $SNR$ values for various optical fibre lengths and meteorological conditions in free space.  From this study we can get a knowledge of the maximum possible achievable distance for both the versions of the protocol and by comparing the results we can evaluate which version of the COW protocol is more feasible for practical implementation.
\subsubsection{Variation of SNR value at different lengths of the optical fibre :}\label{Optical}
One medium for transmitting quantum signals across large distances is Optical Fibre (OF). Single Mode fibre (SMF) is one of the most often used types of fibre among others. Signals experience several losses as they move across the channel (SMF). The presence of impurities in the fibre, light ray scattering inside it and the intense quality of the channel material are the attenuation factors that have the biggest impact on SMF. The attenuation ($\alpha$) in the fibre can be represented as
\begin{eqnarray*}
\alpha = \Big(\frac{10}{L}\Big)\log\Big(\frac{P_{in}}{P_{out}}\Big)
\end{eqnarray*}

 in dB/Km ; where $L$ is the length of fibre, $P_{in}$ is the input signal power of the OF and $P_{out}$ is the output power of the OF. Despite of this loss elements, there are also some other types of losses present in the fibre like dispersion, bending etc.
 
\begin{figure}[h] 
    \centering
    \includegraphics[width = 10 cm, height = 7 cm]{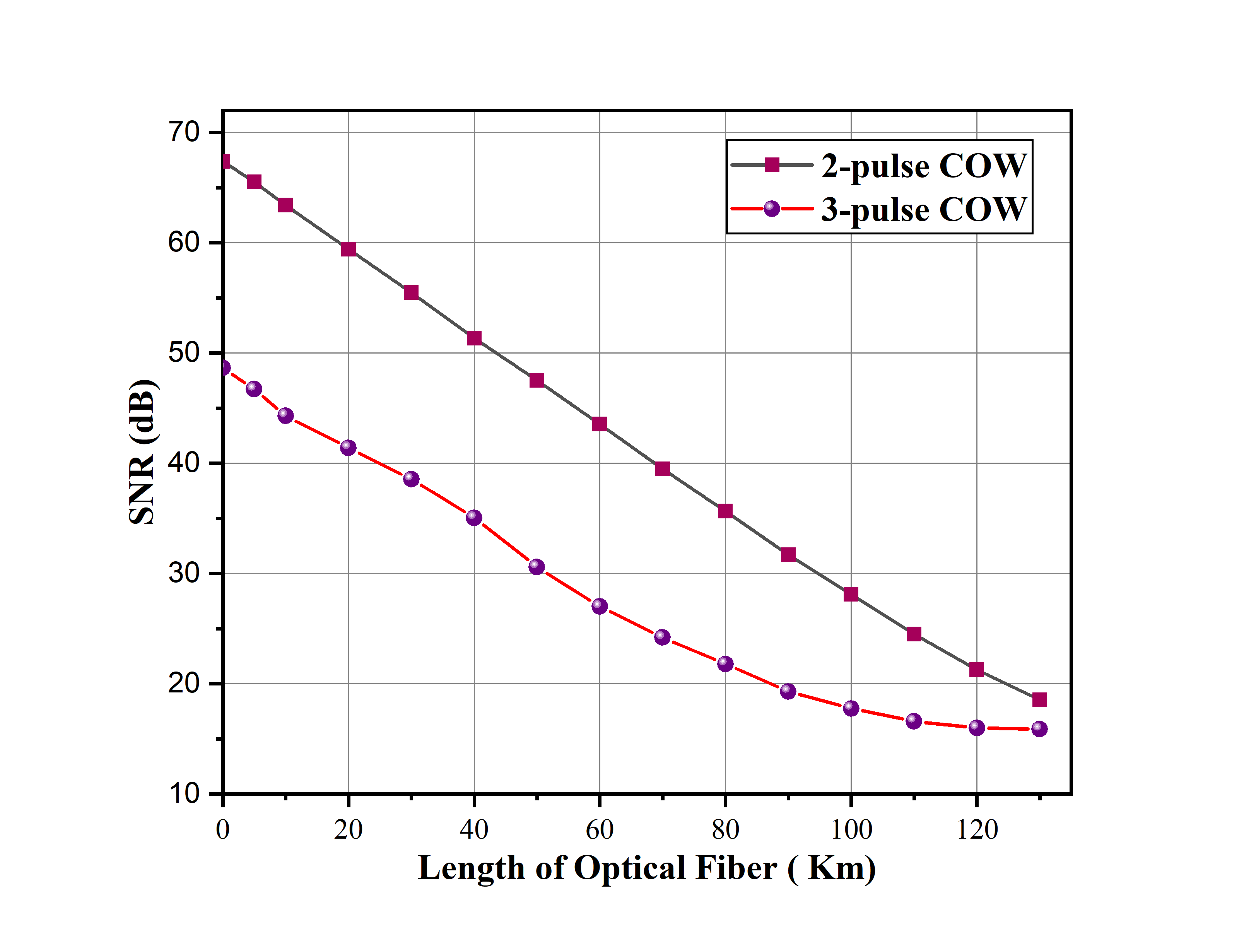}
    \caption{Variation of SNR Values at different Optical fibre lengths.}\label{OF}
\end{figure}

 Now, the variation of $SNR$ values at various fibre distances for both the protocols are displayed in Fig.\ref{OF} and the stipulated parameters of the fibre are mentioned at TABLE \ref{table2}. With reference to the Fig.\ref{OF}, if we only check the $SNR$ value for $70$ Km distance for both the versions then 2-pulses version shows the $SNR$ value of $40$ dB whereas 3-pulses version shows $25$ dB only, Similarly, if we consider the other distances too, then it is pretty visible that at all the points the $SNR$ values are always  higher for 2-pulses version compared to the 3-pulses one.The maximum achievable distance that should be attained is the point at which the power of the signal and noise become equal. By this manner we evaluate the maximum achievable distances for both the protocols which are $120$ Km for 2-pulses version and $70$ Km for 3-pulses version respectively, Considering the $SNR$ values $< 20-25$ dB, we can say that for long distance communication 2-pulses COW version is more effective than the 3-pulses COW QKD protocol.
\subsubsection{Variation of SNR value at different weather conditions in free space :}\label{free}
The primary losses in free space optical (FSO) channel transmission are caused by absorption, scattering, diffraction and turbulence, which are further divided into two categories: a) atmospheric loss and b) geometrical loss. The geometric loss that resulted from the beam broadening during transmission toward the receiving end can be computed using the following expression~\cite{FSO,FSO1,FSO2,FSO3,FSO4}.
\begin{eqnarray*}
\Big [\frac{d_r}{d{_t} + D.L}\Big] 
 \end{eqnarray*}
 where $D$ is the beam divergence, $L$ is the channel length, and $d_r$ and $d_t$ are the diameters of the receiver and transmitter apertures, respectively. 
 
 The atmospheric loss can be explained by  Beer-Lambert's law, i.e., considering  $\exp(-\alpha \cdot L)$ where $\alpha$ is the atmospheric attenuation coefficient, which takes into account the atmospheric medium's absorption and scattering ~\cite{FSO3}. Hence, total loss in FSO can be expressed as follows;
\begin{eqnarray*}
    T = \Big[\frac{d_r}{d{_t} + D.L}\Big]\cdot \exp(-\alpha \cdot L)
\end{eqnarray*}\\
Considering the above mentioned losses in the FSO channels, the attenuation varies due to varying atmospheric turbulence conditions. The specified attenuations are taken from the ~\cite{FSO}, which is also shown in TABLE \ref{tabble1} and the specifications of FSO component are mentioned at TABLE \ref{table2}. Here, we perform and verify the fluctuation of $SNR$ values at different weather conditions at varied free space distances (Fig.~\ref{fig15}).
\begin{figure}[hbt!] 
    \centering
    \includegraphics[width = 6 cm, height = 5.8 cm]{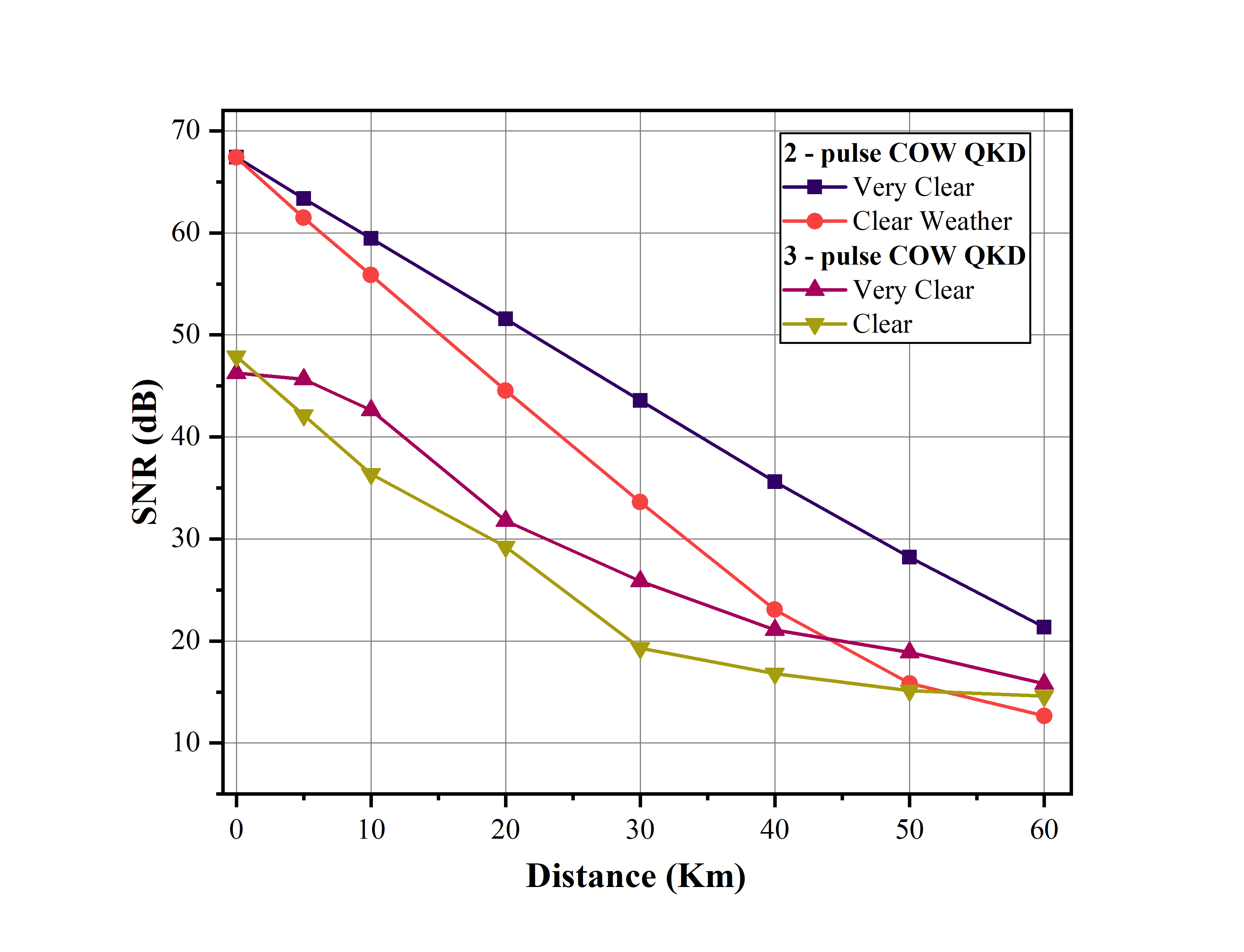}
    \includegraphics[width = 6 cm, height = 5.8 cm]{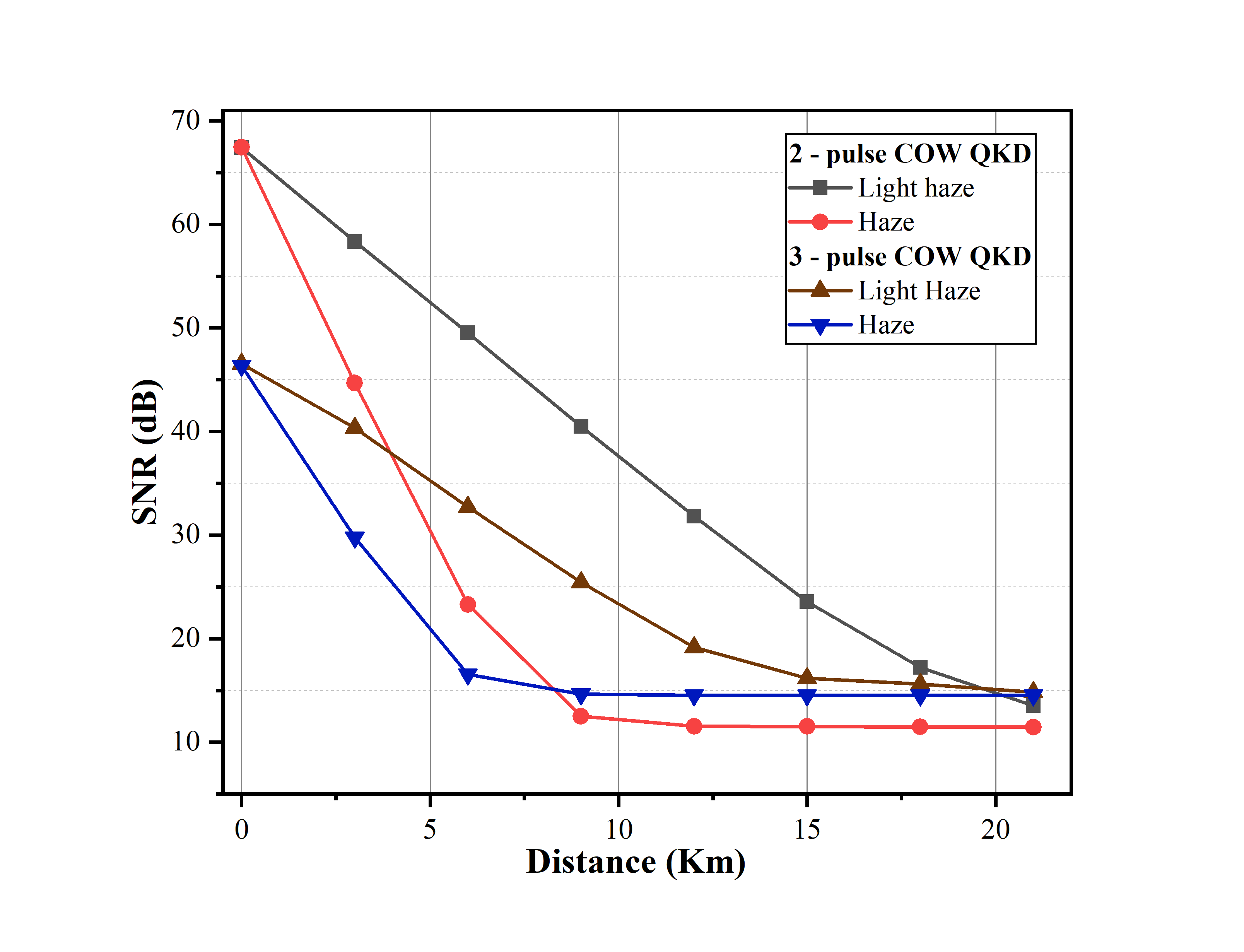}
    (a)$\hspace{5.8cm}$(b)
    \end{figure}
\begin{figure}[hbt!] 
    \centering  
    \includegraphics[width = 6 cm, height = 5.8 cm]{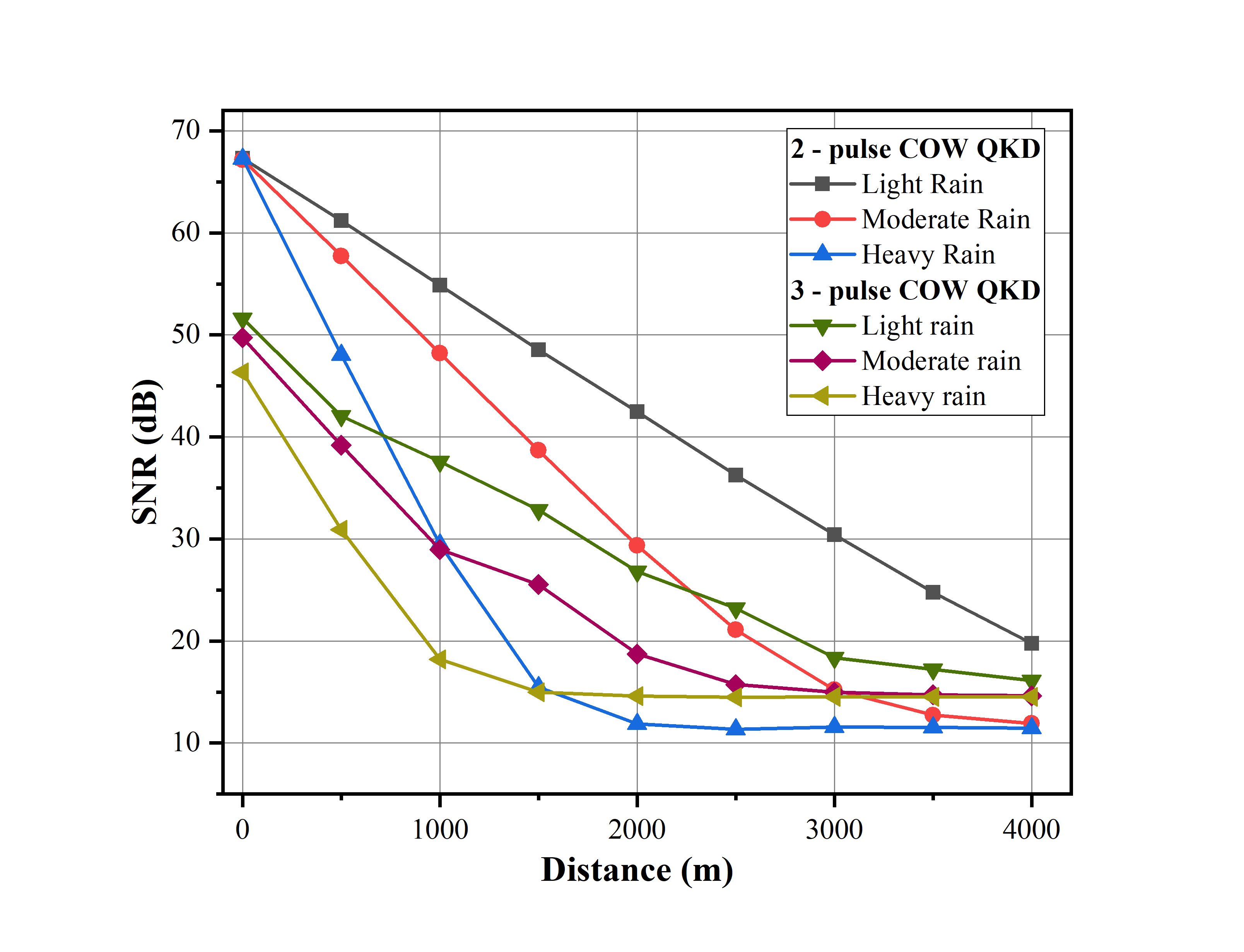}
    \includegraphics[width = 6 cm, height = 5.8 cm]{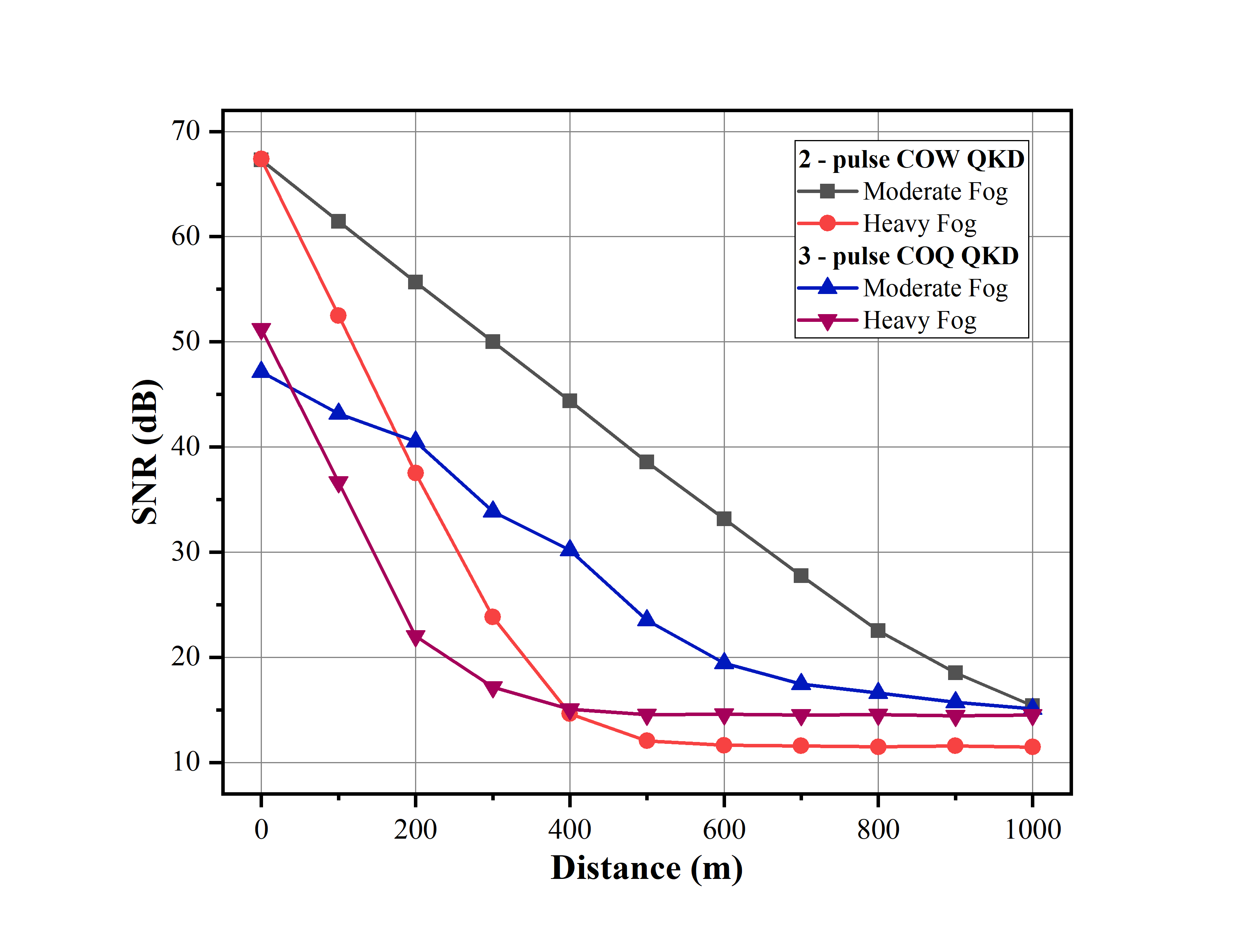}
    (c)$\hspace{5.8cm}$(d)
    \caption{Variation of SNR Values at different weather conditions.}
    \label{fig15}
\end{figure}\\
From the Fig.~\ref{bar}, we can now compare both versions of the protocol at different attenuation levels at different weather circumstances and finally we can say that at all atmospheric turbulences 2-pulses COW is achieving resistance against  long distance compared to 3-pulses one.
\begin{figure}[h] 
    \centering
    \includegraphics[width = 12 cm, height = 8.5 cm]{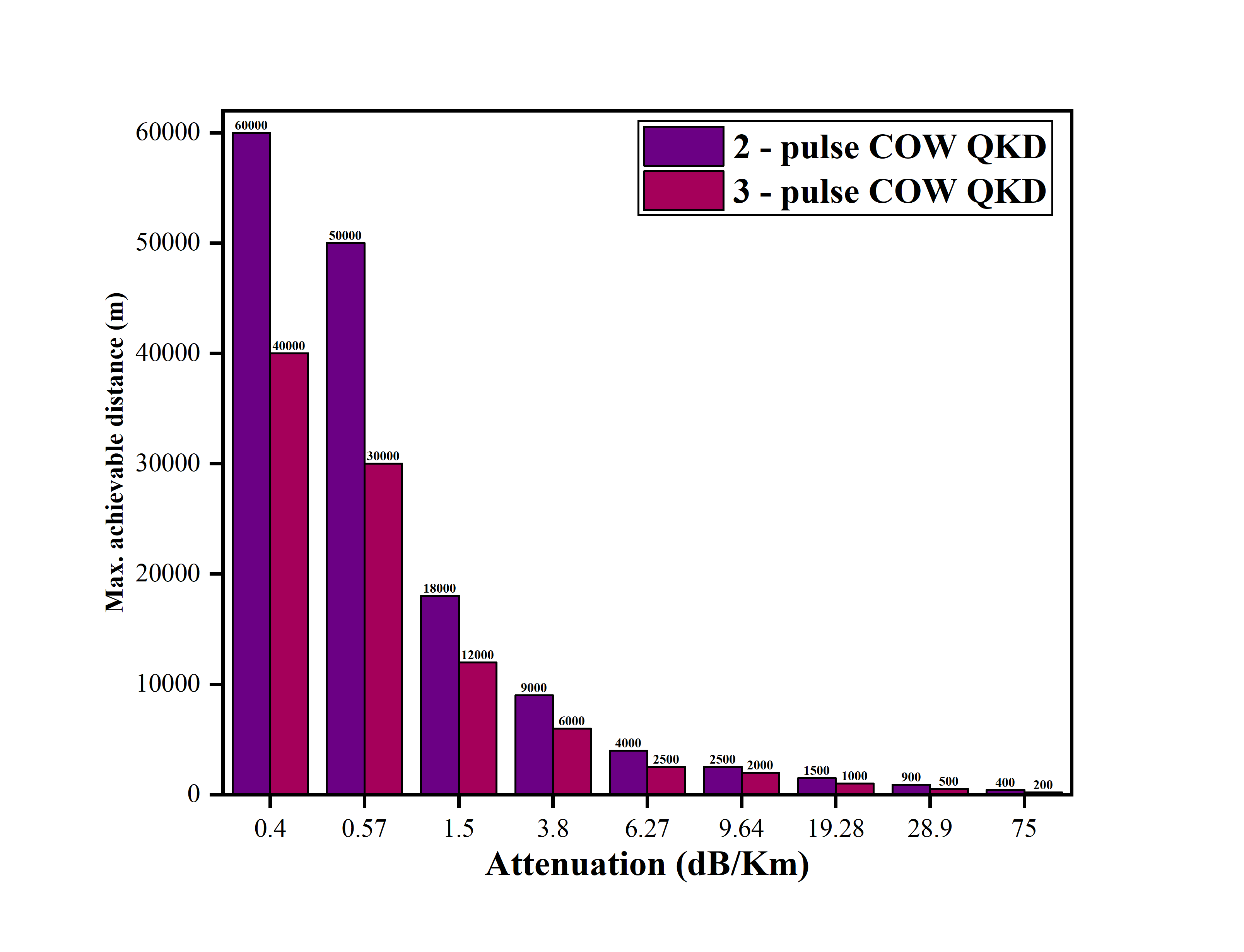}
    \caption{Comparative analysis of SNR values at different atmospheric conditions.}\label{bar}
\end{figure}

\section{Conclusion \& Future Research Work :}\label{conclusion}
We report simulation models of two distinct versions of the COW QKD protocol. In Sec.\ref{SNR}, we perform and analyze the $SNR$ values at wire (optical fibre) and wireless (free space) conditions.  From the results presented in Sec.\ref{Optical} and  comparative analysis [Fig.\ref{bar}] for different FSO conditions for both the versions of the protocol, we can clearly conclude that the states of  2-pulses version is resistance to long distance compared to the 3-pulses version by maintaining proper integrity of the protocol. Due to the extra addition of some optical components in order to incorporate one additional tailor vacuum pulse causes the extra amount of noise creation in the desired signal for 3-pulses version of COW QKD. Although the 3-pulses version of the protocol is unquestionably more secure than the 2-pulses version, it is not appropriate for long-distance communication scenarios. It might be a future research work to make the 3-pulses version long-distance resistance. The components which are required for the protocol's implementation can be predicted based on the parameters we have specified and the configuration we have designed. 
Our future research direction includes the simulations of different cryptanalytic techniques on both the versions of the protocol which might be a sequel of the current work.

\section*{Declaration :}
The authors declare no conflict of interest.
\section*{Data Availability Statement :}
All the data set related to this work will be disclosed upon reasonable request.

\section*{Appendix :}\label{appendix}
The whole security of the protocol is on maintaining the coherence of the states. The coherent state is described as follows;
\begin{center}
\begin{eqnarray*}
 \vert\alpha\rangle = e^{- |\alpha|^{2}/2}
  \sum_{n=1}^{\infty}\frac{\alpha^n}{\sqrt{n!}}\vert n\rangle  
\end{eqnarray*}
\end{center}
where $\alpha$ = Mean number of photons, n = Number of photons.\\
$\Rightarrow$ Expected number of photons in the state $\vert\alpha\rangle$ (using Annihilation (a) and Creation (a$^{\dagger}$) operator);\\
\begin{center}
$\langle n \rangle$ $\equiv$ $\bar{n}$ = $\langle\alpha\vert a^{\dagger}a\vert\alpha\rangle$ = $\langle\alpha\vert a^{\dagger}\alpha\vert\alpha\rangle$ = $\langle\alpha\vert\alpha a^{\dagger}\vert\alpha\rangle$ = $\langle\alpha\vert\alpha\alpha^*\vert\alpha\rangle$ = $\alpha\alpha^*\langle\alpha\vert\alpha\rangle$ = $\vert\alpha\vert^2$\\
\end{center}
$\vspace{0.1cm}$

Using a$\vert\alpha\rangle$ = $\alpha\vert\alpha\rangle$; a$^{\dagger}$$\vert\alpha\rangle$ = $\alpha^*\vert\alpha\rangle$ properties where $\alpha$, $\alpha^*$ are real and complex numbers.
\end{document}